\documentclass[english]{revtex4}
\usepackage[T1]{fontenc}
\usepackage[latin9]{inputenc}
\usepackage[a4paper]{geometry}
\usepackage{color}
\usepackage{float}
\usepackage{amsmath}
\usepackage{graphicx}
\usepackage{amssymb}
\usepackage{esint}

\makeatletter

\providecommand{\tabularnewline}{\\}

\@ifundefined{textcolor}{}
{%
 \definecolor{BLACK}{gray}{0}
 \definecolor{WHITE}{gray}{1}
 \definecolor{RED}{rgb}{1,0,0}
 \definecolor{GREEN}{rgb}{0,1,0}
 \definecolor{BLUE}{rgb}{0,0,1}
 \definecolor{CYAN}{cmyk}{1,0,0,0}
 \definecolor{MAGENTA}{cmyk}{0,1,0,0}
 \definecolor{YELLOW}{cmyk}{0,0,1,0}
 }

\@ifundefined{showcaptionsetup}{}{%
 \PassOptionsToPackage{caption=false}{subfig}}
\usepackage{subfig}
\makeatother

\usepackage{babel}

\begin{document}

\title{Supergraph techniques for $D=3$ $\mathcal{N}=1$ broken supersymmetric
theories}

\author{E. A. Gallegos }

\email{gallegos@fma.if.usp.br}

\author{A. J. da Silva}

\email{ajsilva@fma.if.usp.br}

\address{\textit{Instituto de Física, Universidade de São Paulo,}\\
 \textit{Caixa Postal 66318, 05315-970, São Paulo, SP, Brazil.}
\\
}
\begin{abstract}
We enlarge the usual $D=3$ $\mathcal{N}=1$ supergraph techniques
to include the case of (explicitly or spontaneously) broken supersymmetric
gauge theories. To illustrate the utility of these techniques, we
calculate the two-loop effective potential of the SQED$_{3}$ by using
the tadpole and the vacuum bubble methods. In these methods, to investigate
the possibility of supersymmetry breaking, the superfields must be
shifted by $\theta_{\alpha}$ dependent classical superfields (vacuum
expectation values), what implies in the explicit breakdown of supersymmetry
in the intermediate steps of the calculation. Nevertheless, after
studying the minimum of the resulting effective potential, we find
that supersymmetry is conserved, while gauge symmetry is dynamically
broken, with a mass generated for the gauge superfield.
\end{abstract}
\maketitle

\section{INTRODUCTION}

Supersymmetry (susy), if it exists in nature, must be a broken symmetry
since up to now mass degenerate Bose-Fermi supermultiplets have never
been observed. So, every realistic model must include a mechanism
of susy breakdown \cite{Dine-Mason}. On the other hand, the superfield
formalism developed for exactly supersymmetric theories is a powerful\textcolor{black}{{}
technique} for doing calculations and its possible extension to broken
susy is welcome.

The usual way of studying supersymmetry breakdown is\textcolor{red}{{}
}by treating\textcolor{red}{{} }the breaking terms in the Lagrangian
(of quadratic or of higher number of fields) as interaction vertices
to be incorporated as perturbations into the supersymmetry preserving
theory. Still, in the works \cite{boldo-helayel-neto,helayel-1984}
the superfield formalism for $\mathcal{N}=1$ was enlarged to softly
broken supersymmetric models (in which no quadratic ultraviolet divergences
are triggered by the breaking terms) in 3 and 4 dimensions of space-time,
by treating on an equal footing all bilinear terms. The main difficulty
to overcome in this extension is to calculate the inverse of the kernel
of the bilinear part of the Lagrangian to obtain the superpropagators.

In this paper we will focus on the construction of this extension
for treating spontaneously broken supersymmetric gauge models in three
dimensions. This situation involves bilinear breaking terms of forms
different from that studied in \cite{boldo-helayel-neto}, besides
symmetry breaking monomials with higher number of fields. In fact,
for studying the possibility of spontaneous breaking of susy, we must
translate the fields by their vacuum expectation values, including
$\theta_{\alpha}$ coordinate dependent terms. The kernel of the resulting
bilinear part of the Lagrangian is more general than that in \cite{boldo-helayel-neto}
and by using their operator algebra to obtain the superpropagator
of the spinorial gauge superfield, we learned that it needs a completion.
In effect, in \cite{boldo-helayel-neto}, the authors present an algebra
of six antisymmetric plus six symmetric bi-spinor operators, as a
basis on which the bi-spinor superpropagators (of the spinorial gauge
potential) could be expanded. It is interesting to note that these
twelve operators have a closed algebra, even if they fail as a basis
for the more general form of superpropagators that we find in the
example to be discussed below. We show that two other operators are
required to complete a basis in the more general case.

The paper is organized as follows. In Sect. \ref{sec:algebra-operators}
the Supersymmetric Quantum Electrodynamics in 3D (SQED$_{3}$) is
defined in the superfield language, the algebra of operators needed
to invert the kernel of the bilinear part of the Lagrangian is developed,
and the superpropagators of the shifted SQED$_{3}$ are derived. In
Sect. \ref{sec:Effec-Pot} we compute the zero-, the one- and the
two-loop corrections to the effective potential. The Conclusions,
Sect. \ref{sec:CONCLUDING-REMARKS}, contain some discussions of the
results. Details of the calculations of the effective potential are
given in the Appendices.

\section{THE MODEL AND THE ALGEBRA OF OPERATORS IN 3D\label{sec:algebra-operators}}

In the notation of \cite{sgates-grisaru-rocek-wsiegel} (see also
our Appendix \ref{sec:AppendixA}), the $\mathcal{N}=1$ SQED$_{3}$
is defined by the action\begin{eqnarray}
S & = & \int d^{5}z\left\{ \frac{1}{2}W^{\alpha}W_{\alpha}-\frac{1}{2}\overline{\nabla^{\alpha}\Phi}\nabla_{\alpha}\Phi+M\,\overline{\Phi}\Phi\right\} ,\label{eq:SQED3-Action}\end{eqnarray}
where $\alpha,\,\beta=1,\,2$ are spinorial indices. The UV finiteness
of this model to all loop orders was studied in \cite{Lehum-Adilson}.
The basic elements involved in (\ref{eq:SQED3-Action}) are the complex
(matter) scalar superfield,\begin{equation}
\Phi\left(x,\,\theta\right)=\frac{1}{\sqrt{2}}\left(\Sigma+i\,\Pi\right)=\varphi\left(x\right)+\theta^{\alpha}\psi_{\alpha}\left(x\right)-\theta^{2}F\left(x\right),\end{equation}
where $\Sigma$ and $\Pi$ are real superfields and the component
fields $\varphi$, $\psi_{\alpha}$ and $F$ are respectively a complex
scalar field, a Dirac field and a complex scalar auxiliary field.
The spinor gauge potential is given by\begin{eqnarray}
A_{\alpha}\left(x,\,\theta\right) & = & \chi_{\alpha}\left(x\right)-\theta_{\alpha}B\left(x\right)+i\theta^{\beta}V_{\beta\alpha}\left(x\right)-\theta^{2}\left[2\lambda_{\alpha}\left(x\right)+i\partial_{\alpha\beta}\chi^{\beta}\left(x\right)\right],\end{eqnarray}
where $\chi_{\alpha}$ and $B$ are auxiliary fields, $\lambda_{\alpha}$
is the photino field and $V{}_{a}\equiv(\gamma_{a})^{\alpha\beta}V_{\alpha\beta}$
($a,\, b=0,\,1,\,2$ are Lorentz indices) the 3-vector electromagnetic
potential. The gauge superfield strength is defined as $W_{\alpha}=\frac{1}{2}D^{\beta}D_{\alpha}A_{\beta}$
and has, as one of its component fields, $F_{ab}=\partial_{a}V_{b}-\partial_{b}V_{a}$
, the electromagnetic 3-tensor field strength. 

The susy covariant spinorial derivative is given by $D_{\alpha}=\partial_{\alpha}+i\theta^{\beta}\partial_{\alpha\beta}$
and the susy and gauge covariant derivative by $\nabla_{\alpha}=D_{\alpha}-ieA_{\alpha}$.
The action (\ref{eq:SQED3-Action}) is invariant under the following
gauge transformations:\begin{eqnarray}
\Phi & \rightarrow & \Phi'=e^{ieK}\Phi\nonumber \\
A_{\alpha} & \rightarrow & A'_{\alpha}=A_{\alpha}+D_{\alpha}K,\label{eq:gauge-transf}\end{eqnarray}
where $e$ is the gauge coupling constant and $K\left(x,\,\theta\right)$
is a real scalar superfield. Under these transformations, the electromagnetic
field strength and the covariant derivative go in\[
W_{\alpha}\rightarrow W'_{\alpha}=W_{\alpha}\qquad\qquad\nabla_{\alpha}\rightarrow\nabla'_{\alpha}=e^{ieK}\nabla_{\alpha}e^{-ieK}.\]
 We choose to work with the Lorentz-like gauge fixing term,\begin{equation}
S_{FG}=\int d^{5}z\left(-\frac{1}{4\alpha}\right)D^{\alpha}A_{\alpha}D^{2}D^{\beta}A_{\beta},\label{fg_action}\end{equation}
where $\alpha$ is a dimensionless parameter. With this gauge choice,
the ghosts are free and can be ignored. By adding (\ref{fg_action})
to (\ref{eq:SQED3-Action}), writing $\Phi$ in terms of $\Sigma$
and $\Pi$, and integrating by parts, the full action reads\begin{eqnarray}
S & = & \int d^{5}z\left\{ \frac{1}{2}A_{\alpha}\left[-\frac{1}{2}D^{2}D^{\beta}D^{\alpha}+\frac{1}{2\alpha}D^{2}D^{\alpha}D^{\beta}\right]A_{\beta}+\frac{1}{2}\Sigma\left(D^{2}+M\right)\Sigma+\frac{1}{2}\Pi\left(D^{2}+M\right)\Pi\right.\nonumber \\
 &  & \left.+\frac{e}{2}\left(\Sigma D^{\alpha}\Pi A_{\alpha}-\Pi D^{\alpha}\Sigma A_{\alpha}\right)-\frac{e^{2}}{2}A^{2}\left(\Sigma^{2}+\Pi^{2}\right)\right\} .\label{Action}\end{eqnarray}

In order to compute the effective potential of non-susy theories,
one counts on three popular methods: the Coleman and Weinberg \cite{coleman-weinberg},
the tadpole \cite{weinberg-1973} and the vacuum bubble \cite{jackiw}
methods. In principle, all these methods can be implemented in both
superfields and component fields of supersymmetric gauge theories.
In four dimensions, the one-loop effective potential of the supersymmetric
QED model (along with other two susy models) was evaluated by implementing
the Coleman-Weinberg method in the superfield formalism \cite{Grisaru-Riva-Zanon}.
However, the implementation of this method is cumbersome or even impossible
beyond the one-loop order. On the other hand, the other two methods
are simpler and will be used to compute the effective potential of
the SQED$_{3}$ model. 

To this end, we must shift the superfield $\Sigma$ in (\ref{Action})
by a classical superfield $\sigma(\theta)$: $\Sigma\rightarrow\Sigma+\sigma(\theta)$.
As we want to study the possibility of susy breaking, this classical
field $\sigma(\theta)$ must include a non zero (component) auxiliary
field $\sigma_{2}$, that is, we must consider\begin{equation}
\sigma\left(\theta\right)=\sigma_{1}-\theta^{2}\sigma_{2}.\label{eq:sigma-class}\end{equation}

The resulting expression for the shifted action is

\begin{eqnarray}
S^{\prime}[\sigma_{1},\,\sigma_{2};\,\Sigma,\,\Pi,\, A_{\alpha}] & \equiv & \int d^{5}z\left\{ \frac{1}{2}A_{\alpha}\left[-\frac{1}{2}D^{2}D^{\beta}D^{\alpha}+\frac{1}{2\alpha}D^{2}D^{\alpha}D^{\beta}+\frac{e^{2}}{2}\sigma^{2}\left(\theta\right)C^{\alpha\beta}\right]A_{\beta}\right.\nonumber \\
 &  & +\frac{1}{2}\Pi\left(D^{2}+M\right)\Pi+\xi A^{\alpha}\left[\frac{e}{2}\left(\sigma\left(\theta\right)D_{\alpha}-D_{\alpha}\sigma\left(\theta\right)\right)\right]\Pi+\frac{1}{2}\Sigma\left(D^{2}+M\right)\Sigma\nonumber \\
 &  & +\frac{e}{2}\left(\Sigma D^{\alpha}\Pi A_{\alpha}-\Pi D^{\alpha}\Sigma A_{\alpha}\right)-e^{2}\sigma\left(\theta\right)A^{2}\Sigma-\frac{e^{2}}{2}A^{2}\left(\Sigma^{2}+\Pi^{2}\right)\nonumber \\
 &  & \left.+(D^{2}\sigma+M\sigma)\Sigma+\frac{1}{2}\sigma\left(D^{2}+M\right)\sigma\right\} .\label{eq:shifted-action}\end{eqnarray}
\textcolor{black}{Furthermore we introduced the parameter $\xi$ (to
be made $\xi=1$ at the end of the calculations) in front of the mixing
$\left(A_{\alpha},\,\Pi\right)$ terms to allow to track the effects
of the mixture in the intermediate steps of the calculations.}

The bilinear part of the action, with external source terms added,
reads\begin{eqnarray}
S_{bil} & = & \int d^{5}zd^{5}z^{\prime}\left[\frac{1}{2}A_{\alpha}\left(z\right)\mathcal{O}^{\alpha\beta}\left(z,\, z^{\prime}\right)A_{\beta}\left(z^{\prime}\right)+\frac{1}{2}\Pi\left(z\right)\mathcal{O}\left(z,\, z^{\prime}\right)\Pi\left(z^{\prime}\right)+\xi A_{\alpha}\left(z\right)\mathcal{O}^{\alpha}\left(z,\, z^{\prime}\right)\Pi\left(z^{\prime}\right)\right.\nonumber \\
 &  & \left.+\frac{1}{2}\Sigma\left(z\right)\mathcal{O}\left(z,\, z^{\prime}\right)\Sigma(z^{\prime})+J\left(z\right)\Pi\left(z\right)+\eta^{\alpha}\left(z\right)A_{\alpha}\left(z\right)+G(z)\Sigma(z)\right],\label{APi-square-shifted-action}\end{eqnarray}
where the kernel operators $\mathcal{\mathcal{O}}$ are functions,
not only of the susy covariant operators $D_{\alpha}$ and $\partial_{\alpha\beta}$
(and their square powers), but also of $\theta_{\alpha}$ and $\theta^{2}$:
\begin{subequations}\begin{eqnarray}
\mathcal{O}_{\alpha\beta}\left(z,\, z^{\prime}\right) & = & \mathbf{[}\frac{(1-\alpha)}{2\alpha}i\partial_{\alpha\beta}D^{2}+\frac{1}{2}(-\frac{1+\alpha}{\alpha}\square+e^{2}\sigma_{1}^{2})C_{\alpha\beta}-e^{2}\sigma_{1}\sigma_{2}C_{\alpha\beta}\theta^{2}]\delta^{5}(z-z^{\prime})\label{eq:kernel-ab}\\
\mathcal{O}\left(z,\, z^{\prime}\right) & = & [M+D^{2}]\delta^{5}(z-z^{\prime})\label{eq:kernel-pi}\\
\mathcal{O}_{\alpha}\left(z,\, z^{\prime}\right) & = & [-\frac{e\sigma_{2}}{2}\theta_{\alpha}-\frac{e\sigma_{1}}{2}D_{\alpha}+\frac{e\sigma_{2}}{2}\theta^{2}D_{\alpha}]\delta^{5}(z-z^{\prime}).\label{eq:kernel-api}\end{eqnarray}

\end{subequations}

It must be noted that the mixing between the gauge field and the matter
field represented in $\mathcal{O_{\alpha}}$ can, in general, be avoided
by using an $R_{\xi}$ gauge. This is not true when $\sigma_{2}\neq0.$
In this case the mixture is unavoidable and the use of the Lorentz-like
gauge fixing term has the advantage of having a decoupled ghost sector.

From this action, the superpropagators can be calculated in the usual
way. We start by considering the generating functional $\mathcal{Z}[J,\eta]$
: \begin{equation}
\mathcal{Z}[J,\eta]=\mathcal{N}\int\mathcal{D}\Sigma\mathcal{D}\Pi\mathcal{D}A_{\alpha}\exp\left(i\, S_{bil}\right),\label{APi-generating-functional}\end{equation}
change the superfields by\begin{eqnarray}
\Sigma\left(z\right) & \rightarrow & \Sigma\left(z\right)-\int d^{5}z^{\prime}\Delta_{\Sigma}\left(z,\, z^{\prime}\right)G\left(z^{\prime}\right),\nonumber \\
\Pi\left(z\right) & \rightarrow & \Pi\left(z\right)-\int d^{5}z^{\prime}\{\Delta\left(z,\, z^{\prime}\right)J\left(z^{\prime}\right)+\Delta^{\alpha}\left(z,\, z^{\prime}\right)\eta_{\alpha}\left(z^{\prime}\right)\},\\
A_{\alpha}\left(z\right) & \rightarrow & A_{\alpha}\left(z\right)-\int d^{5}z^{\prime}\{J\left(z^{\prime}\right)\Delta_{\alpha}\left(z^{\prime},\, z\right)+\Delta_{\alpha}^{\hspace{6pt}\beta}\left(z,\, z^{\prime}\right)\eta_{\beta}\left(z^{\prime}\right)\},\nonumber \end{eqnarray}
and determine the superpropagators $\triangle$ by imposing that the
terms which mix fields with currents add to zero. With these conditions,
the integration in the shifted superfields can be carried out, leaving
$\mathcal{Z}[J,\,\eta]$ as a functional of the sources:\begin{eqnarray}
\mathcal{Z}[J,\,\eta] & = & \exp\left[i\int\int d^{5}zd^{5}z^{\prime}\left\{ -\frac{1}{2}\eta^{\alpha}\left(z\right)\Delta_{\alpha}^{\hspace{6pt}\beta}\left(z,\, z^{\prime}\right)\eta_{\beta}\left(z^{\prime}\right)-\frac{1}{2}J\left(z\right)\Delta\left(z,\, z^{\prime}\right)J\left(z^{\prime}\right)\right.\right.\nonumber \\
 &  & \left.\left.-J\left(z\right)\Delta^{\alpha}\left(z,\, z^{\prime}\right)\eta_{\alpha}\left(z'\right)-\frac{1}{2}G\left(z\right)\Delta_{\Sigma}\left(z,\, z^{\prime}\right)G\left(z^{\prime}\right)\right\} \right].\label{eq:zeta}\end{eqnarray}
From this expression, it follows that the superpropagators are given
by \begin{subequations}\begin{eqnarray}
\left\langle T\, A_{\alpha}\left(z\right)A_{\beta}\left(z^{\prime}\right)\right\rangle  & = & i\Delta_{\alpha\beta}\left(z,\, z^{\prime}\right)=i\Theta^{-1}{}_{\alpha\beta}\left(z,\, z^{\prime}\right),\label{eq:invers=0000E3o1}\\
\nonumber \\\left\langle T\,\Pi\left(z\right)\Pi\left(z^{\prime}\right)\right\rangle  & = & i\Delta\left(z,\, z^{\prime}\right)\nonumber \\
 & = & i\mathcal{O}^{-1}\left(z,\, z^{\prime}\right)+i\xi^{2}\int\int_{z_{1},\, z_{2}}\mathcal{O}^{-1}\left(z,\, z_{1}\right)H\left(z_{1},\, z_{2}\right)\mathcal{O}^{-1}\left(z_{2},\, z^{\prime}\right),\\
\nonumber \\\left\langle T\,\Pi\left(z\right)A_{\alpha}\left(z^{\prime}\right)\right\rangle  & = & i\Delta_{\alpha}\left(z,\, z^{\prime}\right)\nonumber \\
 & = & i\xi\int\int_{z_{1},\, z_{2}}\ \mathcal{O}^{-1}\left(z,\, z_{1}\right)\mathcal{O}^{\beta}\left(z_{2},\, z_{1}\right)\Theta^{-1}{}_{\beta\alpha}\left(z_{2},\, z'\right),\\
\nonumber \\\left\langle T\,\Sigma\left(z\right)\Sigma\left(z^{\prime}\right)\right\rangle  & = & i\mathcal{O}^{-1}(z,z')\label{eq:invers=0000E3o4}\end{eqnarray}
\end{subequations}with \begin{eqnarray}
\Theta_{\alpha\beta}\left(z,\, z^{\prime}\right) & = & \mathcal{O}_{\alpha\beta}\left(z,\, z^{\prime}\right)+\xi^{2}Q_{\alpha\beta}\left(z,\, z^{\prime}\right),\nonumber \\
H\left(z,\, z^{\prime}\right) & = & \int\int_{z_{1},z_{2}}\mathcal{O}^{\alpha}\left(z_{1},\, z\right)\Theta^{-1}{}_{\alpha}^{\hspace{6pt}\beta}\left(z_{1},\, z_{2}\right)\mathcal{O}_{\beta}\left(z_{2},\, z^{\prime}\right),\label{eq:invers=0000E3o5}\\
Q_{\alpha\beta}\left(z,\, z^{\prime}\right) & = & \int\int_{z_{1},z_{2}}\mathcal{O}_{\alpha}\left(z,\, z_{1}\right)\mathcal{O}^{-1}\left(z_{1},\, z_{2}\right)\mathcal{O}_{\beta}\left(z^{\prime},\, z_{2}\right).\nonumber \end{eqnarray}

To \textcolor{black}{explicitly} find these superpropagators we develop
the algebra of operators used for calculating the inverse of the matrices
$\mathcal{O}$. Let us begin by considering the scalar sector. Any
scalar operator $\mathcal{\mathcal{O}}=\mathcal{\mathcal{O}}\left(\theta_{\alpha},\, D_{\alpha},\, i\partial_{\alpha\beta}\right)$
can be expanded in terms of six scalar operators,\begin{eqnarray}
\mathcal{O} & = & \sum_{i=0}^{5}p_{i}\, P_{i},\end{eqnarray}
defined in \cite{boldo-helayel-neto} as\begin{equation}
P_{0}=1\text{, \ \ \ \ \ \ \ }P_{1}=D^{2}\text{, \ \ \ \ \ \ \ }P_{2}=\theta^{2},\ \ \ \ \ \ \ P_{3}=\theta^{\alpha}D_{\alpha}\text{,\ \ \ \ \ \ }P_{4}=\theta^{2}D^{2},\text{ \ \ \ \ \ \ }P_{5}=i\partial_{\alpha\beta}\theta^{\alpha}D^{\beta},\label{P.operators}\end{equation}
which form a basis in this sector. The coefficients $p_{i}$ are,
in general, functions of the d'Alembert operator $\square$, the parameters
of the theory (masses, coupling constants, etc.) and of the components
$\sigma_{1}$ and $\sigma_{2}$ of the classical superfield.

The product of the operators $P_{i}$ is presented in Table \ref{tab: scalar-P-multiplication}.
In addition, one has the trivial results $P_{0}P_{i}=P_{i}P_{0}=P_{i}$,
with $i=0,...,5$. 

\begin{table}[H]
\centering{}{\scriptsize }\begin{tabular}{|c|c|c|c|c|c|}
\hline 
 & {\scriptsize $P_{1}$ } & {\scriptsize $P_{2}$ } & {\scriptsize $P_{3}$ } & {\scriptsize $P_{4}$ } & {\scriptsize $P_{5}$ }\tabularnewline
\hline 
{\scriptsize $P_{1}$ } & {\scriptsize $\square$ } & {\scriptsize $-P_{0}+P_{3}+P_{4}$ } & {\scriptsize $2P_{1}+P_{5}$ } & {\scriptsize $-P_{1}+\square P_{2}-P_{5}$ } & {\scriptsize $\square(-2P_{0}+P_{3})$ }\tabularnewline
\hline 
{\scriptsize $P_{2}$ } & {\scriptsize $P_{4}$ } & {\scriptsize $0$ } & {\scriptsize $0$ } & {\scriptsize $0$ } & {\scriptsize $0$ }\tabularnewline
\hline 
{\scriptsize $P_{3}$ } & {\scriptsize $-P_{5}$ } & {\scriptsize $2P_{2}$ } & {\scriptsize $P_{3}-2P_{4}$ } & {\scriptsize $2P_{4}$ } & {\scriptsize $2\square P_{2}+P_{5}$ }\tabularnewline
\hline 
{\scriptsize $P_{4}$ } & {\scriptsize $\square P_{2}$ } & {\scriptsize $-P_{2}$ } & {\scriptsize $2P_{4}$ } & {\scriptsize $-P_{4}$ } & {\scriptsize $-2\square P_{2}$ }\tabularnewline
\hline 
{\scriptsize $P_{5}$ } & {\scriptsize $-\square P_{3}$ } & {\scriptsize $0$ } & {\scriptsize $-2\square P_{2}+P_{5}$ } & {\scriptsize $0$ } & {\scriptsize $\square(P_{3}+2P_{4})$ }\tabularnewline
\hline
\end{tabular}\caption{\label{tab: scalar-P-multiplication}Multiplication table in the scalar
sector.}

\end{table}

Working with this basis, the inversion of $\mathcal{O}$ follows immediately.
Since $\mathcal{O}^{-1}=\sum_{i}\widetilde{p}_{i}P_{i}$ in the basis
$\left\{ P_{i}\right\} $, the requirement $\mathcal{O}^{-1}\mathcal{O}=1$
leads, after using the Table \ref{tab: scalar-P-multiplication},
to a soluble system of six equations for the six unknown coefficients
$\widetilde{p}_{i}$.

For the inversion of $\mathcal{O_{\alpha\beta}}$ we need a basis
of bi-spinor operators. In \cite{boldo-helayel-neto} a {}``basis\textquotedblright{}
of twelve bi-spinor operators,\begin{equation}
R_{i}^{\alpha\beta}=i\partial^{\alpha\beta}P_{i},\ \ \ \ \ \ \ \ \ \, S_{i}^{\alpha\beta}=C^{\alpha\beta}P_{i}\text{,}\end{equation}
was introduced.

According to the authors, any bi-spinor operator may be expanded in
terms of $R_{i}$ and $S_{i}$, that is,\begin{eqnarray}
\mathcal{\mathcal{O}}_{\alpha\beta} & = & \sum_{i=0}^{5}\left(r_{i}\, R_{i,\alpha\beta}+s_{i}\, S_{i,\alpha\beta}\right),\label{eq:false-exp}\end{eqnarray}
 where as before $r_{i}=r_{i}\left(\square,\, c\right)$ and $s_{i}=s_{i}\left(\square,\, c\right)$,
with $c$ labeling all the parameters of the theory. The (closed)
operator algebra obeyed by $R_{i}$ and $S_{i}$ is reproduced in
Table \ref{tab:RS.mult.Tab}. In these tables we have defined $P_{ij}\doteq P_{i}P_{j}$,
where the expansion of the result of the multiplication $P_{i}P_{j}$
in terms of the six $P_{i}$ must be read on Table \ref{tab: scalar-P-multiplication}.

\begin{table}[H]
\noindent \begin{centering}
{\scriptsize }\begin{tabular}{|c|c|c|}
\hline 
 & {\scriptsize $R_{j}$} & {\scriptsize $S_{j}$}\tabularnewline
\hline 
{\scriptsize $R_{i}$} & {\scriptsize $\square S_{ij}$} & {\scriptsize $R_{ij}$}\tabularnewline
\hline 
{\scriptsize $S_{i}$} & {\scriptsize $R_{ij}$} & {\scriptsize $S_{ij}$}\tabularnewline
\hline
\end{tabular}
\par\end{centering}{\scriptsize \par}

\begin{raggedright}
\caption{\label{tab:RS.mult.Tab}Partial multiplication table in the gauge
sector, $(XY)^{\alpha\beta}=X^{\alpha\gamma}Y_{\gamma}^{\hspace{6pt}\beta}$.}

\par\end{raggedright}

\end{table}

Even though (\ref{eq:false-exp}) works for the operators $\mathcal{\mathcal{O}{}_{\alpha\beta}}$
found in \cite{boldo-helayel-neto}, it does not work for the inversion
of the more general form of $\mathcal{\mathcal{O}{}_{\alpha\beta}}$
that we have. It should be noted that any antisymmetric bi-spinor
operator $S^{\alpha\beta}$ has only one independent component and
can always be written as $S^{\alpha\beta}=C^{\alpha\beta}[-\frac{1}{2}S_{\hspace{6pt}\gamma}^{\gamma}]$,
where $S_{\hspace{0.2cm}\gamma}^{\gamma}$ is a scalar operator that
can be expanded in terms of the six $P_{i}$. However, not all symmetric
bi-spinor (which have three independent components) can be written
as a product of $i\partial^{\alpha\beta}=\frac{1}{2}[D^{\alpha}D^{\beta}+D^{\beta}D^{\alpha}]$
by a scalar operator expandable in terms of the six $P_{i}$. In fact,
up to two supercovariant (spinorial) derivatives, one has the independent
symmetric operator\begin{equation}
M^{\alpha\beta}\doteq\theta^{\alpha}D^{\beta}+\theta^{\beta}D^{\alpha}.\label{eq:M.operator}\end{equation}
That $M^{\alpha\beta}$ is independent of the $R_{i}^{\alpha\beta}$
can be seen by explicitly applying $M$, or a linear combination of
the six $R_{i}$ operators, to an arbitrary superfield and verifying
that there is no way of choosing the coefficients of the linear combination
of $R_{i}$ to get the same result. More easily, let us see an example
of inconsistency that appears if we assume that $M$ is a superposition
of the $R_{i}$. Suppose that\begin{equation}
\theta_{\alpha}D_{\beta}+\theta_{\beta}D_{\alpha}=\sum\limits _{i=0}^{5}r_{i}\, R_{i,\alpha\beta}.\label{eq:M.soma.R}\end{equation}
The coefficients $r_{i}$ can be determined by contracting the two
sides with $i\partial^{\alpha\beta}.$ Using the relations $\theta_{\alpha}\theta_{\beta}=-C_{\alpha\beta}\theta^{2}$
and $\partial^{\alpha\beta}\partial_{\beta\gamma}=\delta_{\gamma}^{\alpha}\square$,
along with the above definitions and multiplication tables, this expression
reduces to\begin{equation}
\theta_{\alpha}D_{\beta}+\theta_{\beta}D_{\alpha}=-\frac{1}{\square}R_{5,\alpha\beta}.\label{eq:M.R5}\end{equation}
If we now multiply both sides of (\ref{eq:M.R5}) on the left by $\theta^{\alpha},$
we get the inconsistency\begin{equation}
3\theta^{2}D_{\beta}=\theta^{2}D_{\beta},\end{equation}
showing that the assumption (\ref{eq:M.soma.R}) is incorrect.

Now, by starting with $M^{\alpha\beta}$ we can define six new operators
$M_{i}^{\alpha\beta}\doteq P_{i}M^{\alpha\beta}$ with at most three
spinorial covariant derivatives (in the $\mathcal{N}=1$ superfield
formalism in 3 dimensions, the product of three or more covariant
spinorial derivatives can be reduced to products of two or less spinorial
covariant derivatives $D^{\alpha}$ and the (also susy covariant)
spacetime derivative $i\partial_{\alpha\beta}$). After a little algebraic
work we can see that $M_{i}^{\alpha\beta}=0$, for $i=2,\,3,\,4$
and $5$, and so, the only new operator, aside from $M_{0}^{\alpha\beta}=M^{\alpha\beta}$,
is $M_{1}^{\alpha\beta}=D^{2}M^{\alpha\beta}.$ For convenience, instead
of using $M_{1}$, we will work with\begin{equation}
N^{\alpha\beta}\doteq i\theta^{\alpha}\partial^{\beta\gamma}D_{\gamma}+i\theta^{\beta}\partial^{\alpha\gamma}D_{\gamma}=-M_{1}^{\alpha\beta}+2R_{0}^{\alpha\beta}.\label{N.operator}\end{equation}

The multiplication table of the operators $M$ and $N$ with the twelve
($R$, $S$) ones, that complements the Table (\ref{tab:RS.mult.Tab}),
is shown in the Table \ref{tab:Gauge-Sector}.

\begin{table}[H]
\begin{centering}
\subfloat[]{\begin{centering}
{\scriptsize }\begin{tabular}{|c|c|c|}
\hline 
 & {\scriptsize $N$} & {\scriptsize $M$}\tabularnewline
\hline 
{\scriptsize $S_{0}$} & {\scriptsize $N$} & {\scriptsize $M$}\tabularnewline
\hline 
{\scriptsize $S_{1}$} & {\scriptsize $\ensuremath{-\square M+2R_{1}}$} & {\scriptsize $\ensuremath{-N+2R_{0}}$}\tabularnewline
\hline 
{\scriptsize $S_{2}$} & {\scriptsize $0$} & {\scriptsize $0$}\tabularnewline
\hline 
{\scriptsize $S_{3}$} & {\scriptsize $\ensuremath{N-2R_{4}}$} & {\scriptsize $\ensuremath{M-2R_{2}}$}\tabularnewline
\hline 
{\scriptsize $S_{4}$} & {\scriptsize $\ensuremath{2R_{4}}$} & {\scriptsize $\ensuremath{2R_{2}}$}\tabularnewline
\hline 
{\scriptsize $S_{5}$} & {\scriptsize $\ensuremath{-\square(-M+2R_{2})+2R_{5}}$} & {\scriptsize $N-2R_{3}-2R_{4}$}\tabularnewline
\hline 
{\scriptsize $R_{0}$} & {\scriptsize $\ensuremath{\square(M+S_{3})+R_{5}}$} & {\scriptsize $\ensuremath{N-R_{3}-S_{5}}$}\tabularnewline
\hline 
{\scriptsize $R_{1}$} & {\scriptsize $\ensuremath{-\square(N-R_{3}-2S_{1}-S_{5})}$} & {\scriptsize $\ensuremath{-\square(M-2S_{0}+S_{3})-R_{5}}$}\tabularnewline
\hline 
{\scriptsize $R_{2}$} & {\scriptsize $0$} & {\scriptsize $0$}\tabularnewline
\hline 
{\scriptsize $R_{3}$} & {\scriptsize $\ensuremath{-\square(-M-S_{3}+2S_{4})+R_{5}}$} & {\scriptsize $\ensuremath{N-R_{3}-2\square S_{2}-S_{5}}$}\tabularnewline
\hline 
{\scriptsize $R_{4}$} & {\scriptsize $\ensuremath{2\square S_{4}}$} & {\scriptsize $\ensuremath{2\square S_{2}}$}\tabularnewline
\hline 
{\scriptsize $R_{5}$} & {\scriptsize $\ensuremath{-\square(-Y+2\square S_{2})}$} & {\scriptsize $\ensuremath{-\square(-X+2S_{4})+R_{5}}$}\tabularnewline
\hline 
{\scriptsize $N$} & {\scriptsize $\ensuremath{-\square(-4R_{2}+S_{3}+6S_{4})-2R_{5}}$} & {\scriptsize $\ensuremath{2R_{3}+4R_{4}-6\square S_{2}+S_{5}}$}\tabularnewline
\hline 
{\scriptsize $M$} & {\scriptsize $\ensuremath{-2N-4R_{4}+6\square S_{2}-3S_{5}}$} & {\scriptsize $\ensuremath{-2M-4R_{2}+3S_{3}+6S_{4}}$}\tabularnewline
\hline
\end{tabular}
\par\end{centering}{\scriptsize \par}

}
\par\end{centering}

\begin{centering}
\subfloat[]{\begin{centering}
{\scriptsize }\begin{tabular}{|c|c|c|c|c|c|c|}
\hline 
 & {\scriptsize $\ensuremath{S_{0}}$} & {\scriptsize $\ensuremath{S_{1}}$} & {\scriptsize $\ensuremath{S_{2}}$} & {\scriptsize $\ensuremath{S_{3}}$} & {\scriptsize $\ensuremath{S_{4}}$} & {\scriptsize $\ensuremath{S_{5}}$}\tabularnewline
\hline 
{\scriptsize $N$} & {\scriptsize $N$} & {\scriptsize $\ensuremath{\square M}$} & {\scriptsize $2R_{2}$} & {\scriptsize $\ensuremath{N-2R_{4}}$} & {\scriptsize $\ensuremath{2R_{4}}$} & {\scriptsize $\ensuremath{-\square(M-2R_{2})}$}\tabularnewline
\hline 
{\scriptsize $M$} & {\scriptsize $M$} & {\scriptsize $N$} & {\scriptsize $0$} & {\scriptsize $\ensuremath{M+2R_{2}}$} & {\scriptsize $0$} & {\scriptsize $\ensuremath{-N-2R_{4}}$}\tabularnewline
\hline
\end{tabular}
\par\end{centering}{\scriptsize \par}

}
\par\end{centering}

\begin{centering}
\subfloat[]{\begin{centering}
{\scriptsize }\begin{tabular}{|c|c|c|c|c|c|c|}
\hline 
 & {\scriptsize $R_{0}$} & {\scriptsize $R_{1}$} & {\scriptsize $R_{2}$} & {\scriptsize $R_{3}$} & {\scriptsize $R_{4}$} & {\scriptsize $R_{5}$}\tabularnewline
\hline 
{\scriptsize $N$} & {\scriptsize $\ensuremath{-\square X-R_{5}}$} & {\scriptsize $\ensuremath{-\square Y}$} & {\scriptsize $\ensuremath{2\square S_{2}}$} & {\scriptsize $\ensuremath{-\square(X+2S_{4})-R_{5}}$} & {\scriptsize $2\square S_{4}$} & {\scriptsize $\ensuremath{\square(Y+2\square S_{2})}$}\tabularnewline
\hline 
{\scriptsize $M$} & {\scriptsize $\ensuremath{-Y}$} & {\scriptsize $\ensuremath{-\square X-R_{5}}$} & {\scriptsize $0$} & {\scriptsize $\ensuremath{-Y+2\square S_{2}}$} & {\scriptsize $0$} & {\scriptsize $\ensuremath{-\square(-X+2S_{4})+R_{5}}$}\tabularnewline
\hline
\end{tabular}
\par\end{centering}{\scriptsize \par}

}
\par\end{centering}

\caption{\label{tab:Gauge-Sector}Partial multiplication table in the gauge
sector ($X\doteq M-S_{3}$ and $Y\doteq N-R_{3}+S_{5}$).}

\end{table}

Therefore, the consistent expansion of $\mathcal{O}_{\alpha\beta}$
that replaces (\ref{eq:false-exp}) is given by\begin{eqnarray}
\mathcal{O}_{\alpha\beta} & = & \sum_{i=0}^{5}\left(r_{i}\, R_{i,\alpha\beta}+s_{i}\, S_{i,\alpha\beta}\right)+m\, M_{\alpha\beta}+n\, N_{\alpha\beta},\label{eq:true-exp-gauge}\end{eqnarray}
where the set of fourteen operators $\left\{ R_{i},\, S_{i},\, M,\, N\right\} $
forms a basis in the gauge sector. The inverse operator $\mathcal{O}^{-1}$
is obtained from its definition $\mathcal{O}^{-1,\alpha\beta}\mathcal{O}_{\beta\gamma}=\delta_{\gamma}^{\alpha}$
and the fact that $\mathcal{O}^{-1}$ must have a expansion similar
to that of $\mathcal{\mathcal{O}}$, (\ref{eq:true-exp-gauge}), with
coefficients $\{\widetilde{r}_{i},\,\widetilde{s}_{i},\,\widetilde{m},\,\widetilde{n}\}$
to be determined.

The bilinear mixing terms in (\ref{eq:shifted-action}) give rise
to a spinorial mixing superpropagator $\left\langle T\,\Pi\left(z\right)A_{\alpha}\left(z'\right)\right\rangle $.
As mentioned earlier, this is a consequence of the translation of
the scalar superfield by its vacuum expectation value. So, it is also
convenient to define, for the expansion of $\mathcal{O}_{\alpha}$,
the basis of eight spinorial operators\begin{equation}
\begin{array}{ccccccc}
T_{\alpha}^{1}=\theta_{\alpha} & \hspace{1cm} & T_{\alpha}^{2}=i\partial_{\alpha\beta}\theta^{\beta} & \hspace{1cm} & T_{\alpha}^{3}=\theta_{\alpha}D^{2} & \hspace{1cm} & T_{\alpha}^{4}=i\partial_{\alpha\beta}\theta^{\beta}D^{2}\\
\\T_{\alpha}^{5}=D_{\alpha} &  & T_{\alpha}^{6}=i\partial_{\alpha\beta}D^{\beta} &  & T_{\alpha}^{7}=\theta^{2}D_{\alpha} &  & T_{\alpha}^{8}=i\partial_{\alpha\beta}\theta^{2}D^{\beta}\end{array}\label{eq:mixing-basis}\end{equation}
The results of their multiplications are presented in Table \ref{tab:Mixing-Sector}.

\begin{table}[h]
\begin{centering}
\subfloat[]{\begin{centering}
{\scriptsize }\begin{tabular}{|c|c|c|c|c|}
\hline 
 & {\scriptsize $T_{\beta}^{1}$ } & {\scriptsize $T_{\beta}^{2}$ } & {\scriptsize $T_{\beta}^{3}$ } & {\scriptsize $T_{\beta}^{4}$ }\tabularnewline
\hline 
{\scriptsize $T_{\alpha}^{1}$ } & {\scriptsize $-S_{2}$ } & {\scriptsize $-R_{2}$ } & {\scriptsize $-S_{4}$ } & {\scriptsize $-R_{4}$ }\tabularnewline
\hline 
{\scriptsize $T_{\alpha}^{2}$ } & {\scriptsize $R_{2}$ } & {\scriptsize $\square S_{2}$ } & {\scriptsize $R_{4}$ } & {\scriptsize $\square S_{4}$ }\tabularnewline
\hline 
{\scriptsize $T_{\alpha}^{3}$ } & {\scriptsize $\frac{1}{2}\left(M-S_{3}\right)-S_{4}$ } & {\scriptsize $-R_{4}-\frac{1}{2}S_{5}-\frac{1}{2}N$ } & {\scriptsize $-\square S_{2}+\frac{1}{2}S_{5}+\frac{1}{2}N$ } & {\scriptsize $-\frac{1}{2}\square\left(M-S_{3}+2R_{2}\right)$ }\tabularnewline
\hline 
{\scriptsize $T_{\alpha}^{4}$ } & {\scriptsize $R_{3}+\frac{1}{2}S_{5}-\frac{1}{2}N+R_{4}$ } & {\scriptsize $\frac{1}{2}\square\left(M+S_{3}+2S_{4}\right)+R_{5}$ } & {\scriptsize $-\frac{1}{2}\square\left(M+S_{3}-2R_{2}\right)-R_{5}$ } & {\scriptsize $\square\left(\square S_{2}-R_{3}-\frac{1}{2}S_{5}+\frac{1}{2}N\right)$ }\tabularnewline
\hline 
{\scriptsize $T_{\alpha}^{5}$ } & {\scriptsize $-\frac{1}{2}\left(M+S_{3}\right)+S_{0}$ } & {\scriptsize $R_{0}-R_{3}+\frac{1}{2}S_{5}+\frac{1}{2}N$ } & {\scriptsize $-\frac{1}{2}N+S_{1}+\frac{1}{2}S_{5}$ } & {\scriptsize $\frac{1}{2}\square\left(M-S_{3}\right)+R_{1}+R_{5}$ }\tabularnewline
\hline 
{\scriptsize $T_{\alpha}^{6}$ } & {\scriptsize $-R_{0}-\frac{1}{2}S_{5}+\frac{1}{2}N$ } & {\scriptsize $-\frac{1}{2}\square\left(M-S_{3}+2S_{0}\right)-R_{5}$ } & {\scriptsize $\frac{1}{2}\square\left(M+S_{3}\right)-R_{1}$ } & {\scriptsize $\square\left(R_{3}-\frac{1}{2}S_{5}-\frac{1}{2}N-S_{1}\right)$ }\tabularnewline
\hline 
{\scriptsize $T_{\alpha}^{7}$ } & {\scriptsize $S_{2}$ } & {\scriptsize $R_{2}$ } & {\scriptsize $S_{4}$ } & {\scriptsize $R_{4}$ }\tabularnewline
\hline 
{\scriptsize $T_{\alpha}^{8}$ } & {\scriptsize $-R_{2}$ } & {\scriptsize $-\square S_{2}$ } & {\scriptsize $-R_{4}$ } & {\scriptsize $-\square S_{4}$ }\tabularnewline
\hline
\end{tabular}
\par\end{centering}{\scriptsize \par}

}
\par\end{centering}

\begin{centering}
\subfloat[]{\begin{centering}
{\scriptsize }\begin{tabular}{|c|c|c|c|c|}
\hline 
 & {\scriptsize $T_{\beta}^{5}$ } & {\scriptsize $T_{\beta}^{6}$ } & {\scriptsize $T_{\beta}^{7}$ } & {\scriptsize $T_{\beta}^{8}$ }\tabularnewline
\hline 
{\scriptsize $T_{\alpha}^{1}$ } & {\scriptsize $\frac{1}{2}\left(M-S_{3}\right)$ } & {\scriptsize $-\frac{1}{2}\left(N+S_{5}\right)$ } & {\scriptsize $0$ } & {\scriptsize $0$ }\tabularnewline
\hline 
{\scriptsize $T_{\alpha}^{2}$ } & {\scriptsize $R_{3}+\frac{1}{2}S_{5}-\frac{1}{2}N$ } & {\scriptsize $\frac{1}{2}\square\left(M+S_{3}\right)+R_{5}$ } & {\scriptsize $0$ } & {\scriptsize $0$ }\tabularnewline
\hline 
{\scriptsize $T_{\alpha}^{3}$ } & {\scriptsize $-\frac{1}{2}N-\frac{1}{2}S_{5}$ } & {\scriptsize $\frac{1}{2}\square\left(M-S_{3}\right)$ } & {\scriptsize $-R_{2}+\frac{1}{2}S_{3}+S_{4}-\frac{1}{2}M$ } & {\scriptsize $R_{4}-\square S_{2}+\frac{1}{2}S_{5}+\frac{1}{2}N$ }\tabularnewline
\hline 
{\scriptsize $T_{\alpha}^{4}$ } & {\scriptsize $\frac{1}{2}\square\left(M+S_{3}\right)+R_{5}$ } & {\scriptsize $\square\left(R_{3}+\frac{1}{2}S_{5}-\frac{1}{2}N\right)$ } & {\scriptsize $-R_{3}-\frac{1}{2}S_{5}+\frac{1}{2}N-R_{4}+\square S_{2}$ } & {\scriptsize $\square\left(R_{2}-\frac{1}{2}S_{3}-S_{4}-\frac{1}{2}M\right)-R_{5}$ }\tabularnewline
\hline 
{\scriptsize $T_{\alpha}^{5}$ } & {\scriptsize $R_{0}-S_{1}$ } & {\scriptsize $\square S_{0}-R_{1}$ } & {\scriptsize $R_{2}-\frac{1}{2}S_{3}-S_{4}+\frac{1}{2}M$ } & {\scriptsize $-R_{4}+\square S_{2}-\frac{1}{2}S_{5}-\frac{1}{2}N$ }\tabularnewline
\hline 
{\scriptsize $T_{\alpha}^{6}$ } & {\scriptsize $R_{1}-\square S_{0}$ } & {\scriptsize $\square\left(S_{1}-R_{0}\right)$ } & {\scriptsize $+R_{3}+R_{4}-\square S_{2}+\frac{1}{2}S_{5}-\frac{1}{2}N$ } & {\scriptsize $\square\left(-R_{2}+\frac{1}{2}S_{3}+S_{4}+\frac{1}{2}M\right)+R_{5}$ }\tabularnewline
\hline 
{\scriptsize $T_{\alpha}^{7}$ } & {\scriptsize $R_{2}-S_{4}$ } & {\scriptsize $\square S_{2}-R_{4}$ } & {\scriptsize $0$ } & {\scriptsize $0$ }\tabularnewline
\hline 
{\scriptsize $T_{\alpha}^{8}$ } & {\scriptsize $R_{4}-\square S_{2}$ } & {\scriptsize $\square\left(S_{4}-R_{2}\right)$ } & {\scriptsize $0$ } & {\scriptsize $0$ }\tabularnewline
\hline
\end{tabular}
\par\end{centering}{\scriptsize \par}

}
\par\end{centering}

\caption{\label{tab:Mixing-Sector}Multiplication table in the mixing sector.}

\end{table}

In the momentum space ($i\partial_{\alpha\beta}\rightarrow k_{\alpha\beta}$)
and after an extensive use of these multiplication tables, the superpropagators
(\ref{eq:invers=0000E3o1})-(\ref{eq:invers=0000E3o4}) can be written
as\begin{subequations}\begin{align}
\left\langle T\text{ }A_{\alpha}\left(k,\theta\right)A_{\beta}\left(-k,\theta^{\prime}\right)\right\rangle  & =i\,\left\{ \sum\limits _{i=0}^{5}\left(r_{i}\, R_{i,\alpha\beta}+s_{i}\, S_{i,\alpha\beta}\right)+m\, M_{\alpha\beta}+n\, N_{\alpha\beta}\right\} \delta^{2}\left(\theta-\theta^{\prime}\right),\\[0.1cm]
\left\langle T\text{ }\Pi\left(k,\theta\right)\Pi\left(-k,\theta^{\prime}\right)\right\rangle  & =i\,\left(\sum\limits _{i=0}^{5}a_{i}\, P_{i}\right)\delta^{2}\left(\theta-\theta^{\prime}\right),\\
\left\langle T\,\Pi\left(k,\theta\right)A_{\alpha}\left(-k,\theta^{\prime}\right)\right\rangle  & =i\,\left(\sum\limits _{i=1}^{8}b_{i}\, T_{\alpha}^{i}\right)\delta^{2}\left(\theta-\theta^{\prime}\right),\\
\left\langle T\,\Sigma\left(k,\theta\right)\Sigma\left(-k,\theta^{\prime}\right)\right\rangle  & =i\,\left[c_{0}P_{0}+c_{1}P_{1}\right]\delta^{2}\left(\theta-\theta^{\prime}\right).\end{align}

\label{propagators}\end{subequations}

The coefficients $r_{i}\cdots c_{1}$ are listed in the Appendix \ref{sec:Appedix-B-Prop-Coeff}.
In the rest of the paper we shall study the symmetry properties of
the vacuum of the SQED$_{3}$ model, by calculating the effective
potential up to 2-loops in the perturbation theory. For the 1-loop
calculation, we use the tadpole method \cite{weinberg-1973}, while
for the 2-loop one, we use the vacuum bubble method \cite{jackiw}.
As we will see, susy remains unbroken up to 2-loops, while the internal
$U(1)$ gauge symmetry is broken.

\section{THE EFFECTIVE POTENTIAL UP TO TWO-LOOPS \label{sec:Effec-Pot}}

\subsection{THE CLASSICAL POTENTIAL}

The classical effective action can be read from (\ref{eq:shifted-action}).
The terms depend only on the classical field $\sigma$ are\[
\Gamma_{cl}=\int d^{5}z\frac{1}{2}\sigma\left(D^{2}+M\right)\sigma\equiv-\int d^{3}xU_{cl}(\sigma_{1,}\sigma_{2}),\]
where the second equality defines the Classical Potential. After integrating
in the $\theta$ variables we get $U_{cl}\left(\sigma_{1},\sigma_{2}\right)=-\frac{1}{2}\sigma_{2}^{2}-M\sigma_{1}\sigma_{2}.$
The classical potential can also be obtained by integrating the tree-level
$\Sigma$ supertadpole (\ref{eq:shifted-action}):\begin{equation}
\Gamma_{cl}^{(\Sigma)}=\int d^{5}z\Sigma(D^{2}\sigma+M\sigma)\label{treelevel_tadpole_action}\end{equation}
where, in component fields, $\Sigma\doteq\Sigma_{1}\left(x\right)+\theta^{\alpha}\Psi_{\alpha}\left(x\right)-\Sigma_{2}\left(x\right)\theta^{2}.$
Starting from this tadpole we have two alternatives for computing
the classical potential. We can work in the superfield approach and
adopt the superfield Miller's recipe \cite{millerB228-1983,millerB229-1983}
or we can jump to the component approach. We choose the last option
because it is simpler in the calculations at 1-loop level (next section).
Substituting $\mbox{\ensuremath{\Sigma}}$ in terms of its component
fields in (\ref{treelevel_tadpole_action}) and integrating in $\theta$,
we obtain

\[
\Gamma_{cl}^{(\Sigma)}=\int d^{3}x[M\sigma_{2}\Sigma_{1}(x)+\left(M\sigma_{1}+\sigma_{2}\right)\Sigma_{2}(x)].\]

From this expression we can easily recognize the tree-level $\Sigma_{1}(\Sigma_{2})$
tadpoles and set up the tadpole equations:\begin{eqnarray}
\frac{\partial U_{cl}}{\partial\sigma_{1}} & = & -M\sigma_{2}\\
\frac{\partial U_{cl}}{\partial\sigma_{2}} & = & -\left(M\sigma_{1}+\sigma_{2}\right).\end{eqnarray}

By integrating these equations, we get, as before,\begin{equation}
U_{cl}\left(\sigma_{1},\sigma_{2}\right)=-\frac{1}{2}\sigma_{2}^{2}-M\sigma_{1}\sigma_{2}.\label{classical-potential}\end{equation}

\subsection{ONE- AND TWO-LOOPS CONTRIBUTIONS TO THE EFFECTIVE POTENTIAL}

Having determined the explicit form of the shifted superpropagators,
we are ready to compute the one- and the two-loops contributions to
the effective potential. Since the coefficients of the superpropagators
are merely functions of $k^{2}$ (and of the parameters of the shifted
theory) we do hide their intricate structures in the intermediate
stages of the computations. This is possible because the Grassmann
calculus needed to reduce the $\theta$ integrations to a single $\theta$
integration involves only $(\theta_{\alpha},\, D_{\alpha},\, k_{\alpha\beta})$
manipulations.

\subsubsection{One-loop contribution}

\begin{figure}
\noindent \begin{centering}
\includegraphics[scale=0.4]{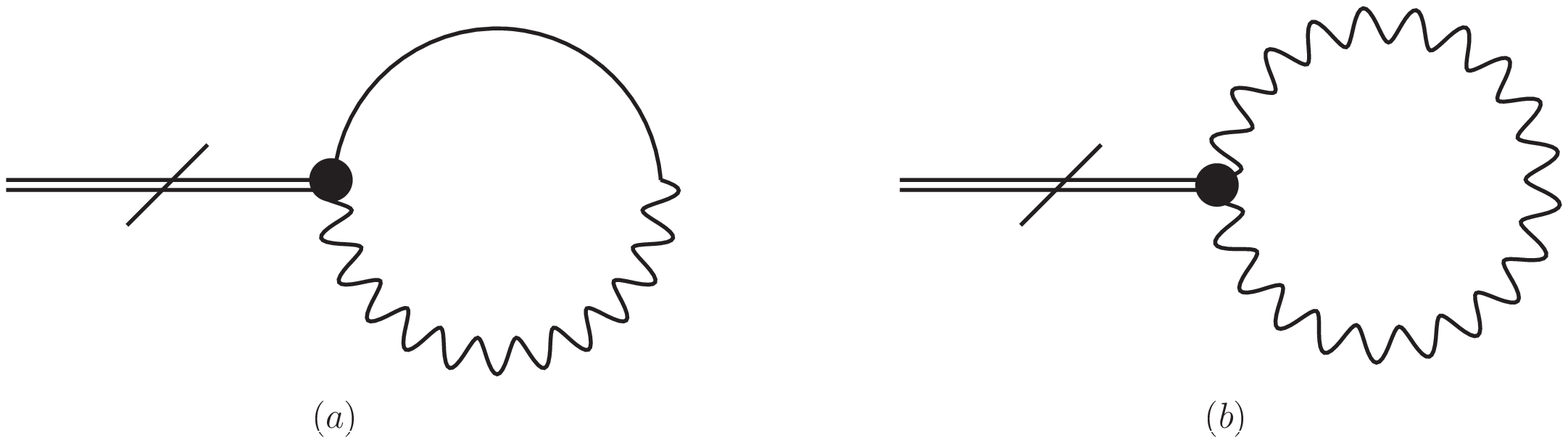}
\par\end{centering}

\raggedright{}\caption{\label{fig:One-loop}One-loop contributions to the $\Sigma$ tadpole
of the shifted SQED$_{3}$. Double-solid line represents the $\Sigma$
scalar superpropagator, solid-wavy line the $\left\langle \Pi A\right\rangle $
mixed superpropagator, and wavy line the gauge superpropagator.}

\end{figure}

At the one-loop order, we use the tadpole method. Figure \ref{fig:One-loop}
shows the two contributions to the tadpoles. Their corresponding integrals
are\begin{eqnarray}
\Gamma_{1} & = & \int d^{2}\theta\int\frac{d^{3}k}{\left(2\pi\right)^{3}}\left[e\left\langle D^{\alpha}\Pi\left(k,\theta\right)A_{\alpha}\left(-k,\theta\right)\right\rangle +\frac{e}{2}\left\langle \Pi\left(k,\theta\right)D^{\alpha}A_{\alpha}\left(-k,\theta\right)\right\rangle \right.\nonumber \\
 &  & \left.-\frac{e^{2}}{2}\sigma\left(\theta\right)\left\langle A^{\alpha}\left(k,\theta\right)A_{\alpha}\left(-k,\theta\right)\right\rangle \right]\int d^{3}x\Sigma(x,\theta)\end{eqnarray}

As discussed in \cite{burgess} and reproduced in our previous paper
\cite{gallegos}, to study the possibility of susy breaking it is
enough to calculate the radiative corrections to the effective potential
up to linear dependence in $\sigma_{2}$. So, in the following we
will restrict our calculations to this approximation. Inserting the
superpropagators (\ref{propagators}) and integrating by parts, one
obtains\begin{eqnarray}
\Gamma_{1} & = & i\int d^{2}\theta\int\frac{d^{3}k}{\left(2\pi\right)^{3}}\left[\left(eb_{5}\left(k\right)-2eb_{3}\left(k\right)+e^{2}\sigma_{1}s_{1}\left(k\right)\right)+\right.\nonumber \\
 &  & \left.\left(eb_{7}\left(k\right)-e^{2}\sigma_{2}s_{1}\left(k\right)+e^{2}\sigma_{1}s_{4}\left(k\right)\right)\theta^{2}\right]\int d^{3}x\Sigma(x,\theta),\end{eqnarray}
which, after integrating in the $\theta$ variables, gives

\begin{eqnarray}
\Gamma_{1} & = & i\int\frac{d^{3}k}{\left(2\pi\right)^{3}}\left[\left(eb_{5}\left(k\right)-2eb_{3}\left(k\right)+e^{2}\sigma_{1}s_{1}\left(k\right)\right)\int d^{3}x\Sigma_{2}(x)\right.\nonumber \\
 &  & \left.-\left(eb_{7}\left(k\right)-e^{2}\sigma_{2}s_{1}\left(k\right)+e^{2}\sigma_{1}s_{4}\left(k\right)\right)\int d^{3}x\Sigma_{1}(x)\right].\end{eqnarray}

From this expression we can directly read the tadpole equations for
the components $\Sigma_{1}$ and $\Sigma_{2}$:\begin{eqnarray}
\frac{\partial U_{1}}{\partial\sigma_{1}} & = & i\int\frac{d^{3}k}{\left(2\pi\right)^{3}}\left[eb_{7}\left(k\right)-e^{2}\sigma_{2}s_{1}\left(k\right)+e^{2}\sigma_{1}s_{4}\left(k\right)\right]\\
\frac{\partial U_{1}}{\partial\sigma_{2}} & = & i\int\frac{d^{3}k}{\left(2\pi\right)^{3}}\left[-eb_{5}\left(k\right)+2eb_{3}\left(k\right)-e^{2}\sigma_{1}s_{1}\left(k\right)\right].\end{eqnarray}

The coefficients $b_{i}$ and $s_{i},$ which are functions of $\sigma_{1}$
and $\sigma_{2},$ are given up to a linear dependence in $\sigma_{2}$
and in $\alpha$ (this last restriction is for simplicity of calculation)
in the Appendix \ref{sec:Appedix-B-Prop-Coeff}. Solving this pair
of equations (in the $\alpha$- and $\sigma_{2}$-linear approximation),
we get\begin{eqnarray}
U_{1}\left(\sigma_{1},\sigma_{2}\right) & = & \frac{\xi\alpha e^{2}M\sigma_{1}\sigma_{2}}{2}\, i^{2}\int\frac{d^{3}k}{\left(2\pi\right)^{3}}\frac{1}{k^{2}\left(k^{2}+M^{2}\right)}\nonumber \\
 & = & -\frac{\alpha\xi}{8\pi}e^{2}\sigma_{1}\sigma_{2},\label{eq:one-loop-potential}\end{eqnarray}
a result that depends on the gauge parameter $\alpha$ and is zero
in the Landau gauge ($\alpha=0$). From (\ref{eq:one-loop-potential})
we see that neither susy nor gauge symmetry are dynamically broken
at one-loop order. This outcome has already been obtained long ago
in \cite{burgess}. 

We should emphasize that the gauge dependence of the effective potential
is a well known fact \cite{dolan-jackiw-1974,nielsen-1975,aitchison-fraser-1984,Johnston-1985}.
In spite of this fact, the Nielsen Identities show that the value
of the potential at its minimum and the values of the generated masses
are independent of the gauge parameter $\alpha$. So, the conclusions
about breakdown of symmetries, got from the analysis of the minimum
of the effective potential, are in fact gauge independent. Let us
now extend our study to the two-loop level.

\subsubsection{Two-loop contributions}

\begin{figure}
\begin{centering}
\includegraphics[scale=0.4]{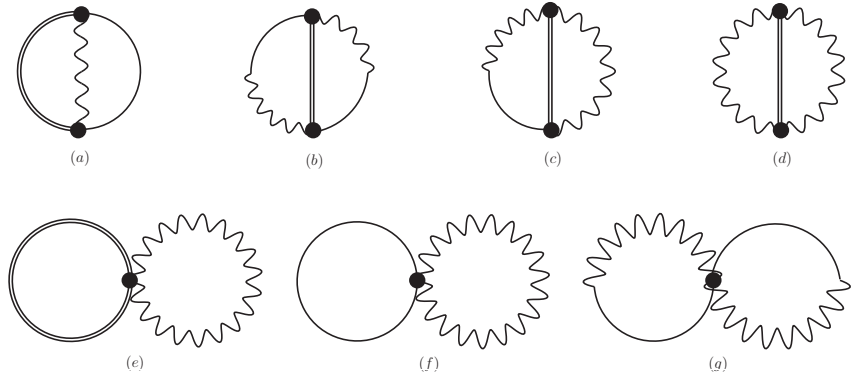}
\par\end{centering}

\raggedright{}\caption{\label{fig:Two-loop}Two-loop vacuum bubbles for the shifted SQED$_{3}$.
Solid lines stand for $\Pi$ scalar superpropagators.}

\end{figure}

At this order we use the vacuum bubble method \cite{jackiw}. The
seven supergraphs that contribute to the effective potential are depicted
in Figure \ref{fig:Two-loop}. Nevertheless, once the integration
over the $\theta$ variables have been carried out \cite{Ferrari},
only the diagrams 2(a), 2(c) and 2(d) survive in our approach (linear
contribution in $\alpha$ and $\sigma_{2}$). The corresponding integrals
are shown in the Appendix \ref{sec:Appendix-C-Two-Loop-Calculations}.
The integrations over the internal momenta were done (with dimensional
regularization) using the results in \cite{Tan-Tekin-Hosotani,Dias-Adilson}.
The result exhibits the following structure\begin{equation}
U_{2}\left(\sigma_{1},\,\sigma_{2}\right)=\sigma_{1}\sigma_{2}\left[F\left(m_{1}^{2},M^{2}\right)+\alpha G\left(m_{1}^{2},M^{2}\right)\right]e^{4},\label{eq:2-loop-effpot-1}\end{equation}
where $m_{1}^{2}=e^{2}\sigma_{1}^{2}/2.$ However, for convenience
of the analysis of the minimum of the effective potential, we rewrite
this result in the form\begin{equation}
U_{2}\left(\sigma_{1},\,\sigma_{2}\right)=\sigma_{2}\left[f\left(\frac{e\sigma_{1}}{M}\right)+\alpha g\left(\frac{e\sigma_{1}}{M}\right)\right]\frac{e^{3}}{64\pi^{2}},\label{eq:2-loop-effpot-2}\end{equation}
where the finite functions $f(x)$ and $g(x)$ are given by\begin{eqnarray}
f(x) & = & \frac{1}{x}+\frac{1}{\sqrt{2}}\frac{1}{1+x/(2\sqrt{2})}-\frac{\sqrt{2}}{1+\sqrt{2}x}+\frac{2}{_{x^{3}}}\ln\left[\frac{1+\sqrt{2}x}{\left(1+\frac{x}{\sqrt{2}}\right)^{2}}\right]\end{eqnarray}
and\begin{eqnarray}
g(x) & = & -\frac{2x}{\left(2+\sqrt{2}x\right)^{2}}+\xi\left(-\frac{2\sqrt{2}x^{2}}{x^{2}+3\sqrt{2}x+4}+4x\ln\left[\frac{2\sqrt{2}+x}{\sqrt{2}+x}\right]\right)\nonumber \\
 &  & +\xi^{2}\left(-\frac{4x\left(6+\sqrt{2}x\right)}{\left(4+\sqrt{2}x\right)^{2}}+x\ln\left[\frac{2\sqrt{2}+x}{\sqrt{2}+x}\right]\right),\end{eqnarray}
At this point a remark is in order. Even if susy or gauge symmetry
is broken, a phase (rotational) symmetry is preserved \cite{coleman-weinberg}
in the effective potential. In our variables $\Sigma$ and $\Pi$
(real and imaginary components of the superfield $\Phi)$, the rotational
gauge symmetry can be recovered by the substitutions $\sigma_{1}^{2}\rightarrow\sigma_{1}^{2}+\pi_{1}^{2}$,
$\sigma_{2}^{2}\rightarrow\sigma_{2}^{2}+\pi_{2}^{2}$ and $\sigma_{1}\sigma_{2}\rightarrow\sigma_{1}\sigma_{2}+\pi_{1}\pi_{2}$
in the results (\ref{classical-potential}), (\ref{eq:one-loop-potential})
and (\ref{eq:2-loop-effpot-1}). Here $\pi_{1}$ and $\pi_{2}$ are
the components of the translation $\pi=\pi_{1}-\theta^{2}\pi_{2}$
in the field $\Pi$ (which for simplicity we did not considered).
This symmetry is not evident when the two-loop contribution to the
effective potential is written in the form (\ref{eq:2-loop-effpot-2}). 

By collecting the results of zero-, one- and two-loops, and by choosing,
for simplicity, the Landau gauge ($\alpha=0$), the effective potential
turns out\begin{eqnarray}
U_{eff}\left(\sigma_{1},\sigma_{2}\right) & = & -\frac{1+\delta z}{2}\sigma_{2}^{2}-\left(M+\delta M\right)\sigma_{1}\sigma_{2}+\frac{e^{3}}{64\pi^{2}}f\left(\frac{e\sigma_{1}}{M}\right)\sigma_{2\text{ }}.\label{eq:pot-eff}\end{eqnarray}

In this result we have introduced two counterterms: the matter field
wave function renormalization counterterm $\delta z$ and the mass
renormalization counterterm $\delta M.$ They must be fixed by the
renormalization prescriptions on the effective potential. It must
be noted that both one- and two-loop contributions (in dimensional
regularization) are finite. Despite its appearance, $f(x)$ is a finite
monotonically decreasing function,\begin{eqnarray}
f\left(x\right) & = & \frac{1}{\sqrt{2}}-\frac{7}{8\sqrt{2}}x^{2}+\frac{133}{96}x^{3}-\frac{223}{64\sqrt{2}}x^{4}+\mathcal{O}\left(x^{5}\right)\text{ \ \ \ \ \ \ for \ }x\ll1,\end{eqnarray}
running from $f(x=0)=1/\sqrt{2}$ to $f(x=\infty)=0.$

As the radiative corrections are finite, their effects in the redefinition
of the mass and the wave function normalization are finite and we
can adopt for convenience a {}``minimal subtraction renormalization
prescription\textquotedblright{}: $\delta z=\delta M=0$, resulting
that, up to two loops, the renormalized effective potential is given
by\begin{equation}
U_{eff}\left(\sigma_{1},\sigma_{2}\right)=-\frac{1}{2}\sigma_{2}^{2}-\sigma_{2}\left[x-\left(\frac{e^{2}}{8\pi M}\right)^{2}f(x)\right]\frac{M^{2}}{e}.\label{eq:effpot-1}\end{equation}
where $x=e\sigma_{1}/M.$

From the Euler Lagrange equation for $\sigma_{2}$, that is, $\partial U_{eff}\left(\sigma_{1},\sigma_{2}\right)/\partial\sigma_{2}=0$,
we get\begin{equation}
\sigma_{2}=\left[\left(\frac{e^{2}}{8\pi M}\right)^{2}f\left(\frac{e\sigma_{1}}{M}\right)-x\right]\frac{M^{2}}{e}.\label{eq:sigma2-1}\end{equation}
 After inserting this result into the expression for $U{}_{eff}(\sigma_{1},\,\sigma_{2})$,
one obtains\begin{equation}
U_{eff}=\frac{M^{4}}{2e^{2}}\left[x-\left(\frac{e^{2}}{8\pi M}\right)^{2}f(x)\right]^{2},\label{eq:effpot-3}\end{equation}
which satisfies $U{}_{eff}\geq0.$ Its minimum $\left(U{}_{eff}=0\right)$
occurs for\begin{equation}
\frac{e\sigma_{1}}{M}=\left[\frac{e^{2}}{8\pi M}\right]_{\text{ }}^{2}f\left(\frac{e\sigma_{1}}{M}\right),\label{eq:solution}\end{equation}
which also implies $\sigma_{2}=0.$ In perturbation expansion, by
hypothesis $e^{2}/8\pi M\ll1$, and the equation (\ref{eq:solution})
has the solution $\sigma_{1}\cong e^{3}/\sqrt{2}64\pi^{2}M\neq0$.
In short, the minimum of the effective potential is $U{}_{eff}=0$
and occurs at $\sigma_{2}=0$ and $\sigma_{1}\neq0$. This result
means \cite{burgess,gallegos} that supersymmetry is preserved, but
the gauge symmetry is dynamically broken, with a mass $m_{1}=(\frac{e^{2}}{8\pi M})^{2}\frac{M}{2}\neq0$
generated for the gauge superfield. By only making the shift $\sigma=\sigma_{1}$,
in paper \cite{Lehum}, the breakdown of the gauge symmetry was studied
with a similar conclusion. As a result of these corrections the gauge
superpropagator (\ref{eq:prop-gauge}) is given by\[
\left\langle T\text{ }A_{\alpha}\left(k,\theta\right)A_{\beta}\left(-k,\theta^{\prime}\right)\right\rangle =\frac{i}{2\left(k^{2}+m_{1}^{2}\right)}\left[-C_{\alpha\beta}+\frac{k_{\alpha\beta}D^{2}}{k^{2}}\right]\delta^{2}\left(\theta-\theta^{\prime}\right).\]

By the component decomposition of $A_{\alpha}$, presented in the
Appendix \ref{sec:AppendixA}, we obtain for the component field propagators\begin{eqnarray*}
\left\langle T\,\chi_{\alpha}\left(k\right)\chi_{\beta}\left(-k\right)\right\rangle  & = & \frac{i~k_{\alpha\beta}}{2k^{2}\left(k^{2}+m_{1}^{2}\right)}\\
\left\langle T\,\lambda_{\alpha}\left(k\right)\lambda_{\beta}\left(-k\right)\right\rangle  & = & -\frac{i~k_{\alpha\beta}}{2\left(k^{2}+m_{1}^{2}\right)}\\
\left\langle T\,\chi_{\alpha}\left(k\right)\lambda_{\beta}\left(-k\right)\right\rangle  & = & -\frac{i~C_{\alpha\beta}}{2\left(k^{2}+m_{1}^{2}\right)}\\
\left\langle T\, V_{a}\left(k\right)V_{b}\left(-k\right)\right\rangle  & = & -\frac{2i}{k^{2}+m_{1}^{2}}\left(\eta_{ab}-\frac{k_{a}k_{b}}{k^{2}}\right)\ \ \ a,b=0,1,2\\
\left\langle T\, B\left(k\right)B\left(-k\right)\right\rangle  & = & 0\,.\end{eqnarray*}
where $V_{a}\equiv(\gamma_{a})^{\alpha\beta}V_{\alpha\beta}$ is the
3-vector electromagnetic potential.

\section{CONCLUDING REMARKS \label{sec:CONCLUDING-REMARKS}}

In this paper we developed the algebra of spinorial operators involved
in the calculation of the superpropagators for gauge and matter field
models in 3D. This algebra is useful in the presence of shifts of
the superfields by $\theta$ spinorial dependent expectation values,
as needed to calculate the effective potential to study the possibility
of dynamical supersymmetry breakdown. As an example, this algebra
is applied in the calculation of the superpropagators of the supersymmetric
quantum electrodynamics SQED$_{3}.$ The shift of superfields with
such a $\theta$ dependent part implies in bilinear mixing of the
gauge and matter fields that cannot be eliminated by using an $R_{\xi}$
gauge fixing term. The inversion of the quadratic part of the Lagrangian
results very arduous in component or in the superfield formalism.
The use of this algebra systematizes the calculation of the superpropagators,
and it is helpful in the calculation of the superpropagators of any
$\mathcal{N}=1$ supersymmetric model in 3D. The effective potential
for SQED$_{3}$ is calculated up to two loops, with the conclusion
that supersymmetry is preserved, but gauge symmetry is dynamically
broken with the generation of mass for the gauge superfield.

\begin{center}
\textbf{ACKNOWLEDGEMENTS} 
\par\end{center}

This work was partially supported by the Conselho Nacional de Desenvolvimento
Cientifico e Tecnológico (CNPq) of Brazil, and Fundação de Amparo
à Pesquisa do Estado de São Paulo (FAPESP). The authors would like
to thank A. F. Ferrari for the implementation of the SusyMath package
to the case of explicit broken supersymmetric theories in 3D.

\appendix

\section{THE SUPERFIELD EXPANSIONS \label{sec:AppendixA}}

In component fields the matter and gauge superfields can be written
as \cite{sgates-grisaru-rocek-wsiegel}: \begin{subequations}\begin{eqnarray}
\Sigma\left(x,\theta\right) & = & \Sigma_{1}\left(x\right)+\theta^{\alpha}\Psi_{\alpha}-\theta^{2}\Sigma_{2}\left(x\right),\text{ \ \ \ \ }\alpha,\beta=1,2\\
\Pi\left(x,\theta\right) & = & \Pi_{1}\left(x\right)+\theta^{\alpha}\Xi_{\alpha}-\theta^{2}\Pi_{2}\left(x\right)\\
A_{\alpha}\left(x,\theta\right) & = & \chi_{\alpha}\left(x\right)-\theta_{\alpha}B\left(x\right)+i\theta^{\beta}V_{\beta\alpha}\left(x\right)-\theta^{2}\left[2\lambda_{\alpha}\left(x\right)+i\partial_{\alpha\beta}\chi^{\beta}\left(x\right)\right]\\
W_{\alpha}\left(x,\theta\right) & = & \lambda_{\alpha}\left(x\right)+\theta^{\beta}f_{\beta\alpha}\left(x\right)+\theta^{2}i\partial_{\alpha\beta}\lambda^{\beta}\left(x\right)\end{eqnarray}
\end{subequations}with $f_{\alpha\beta}\left(x\right)=-\frac{1}{2}\left(\partial_{\alpha}^{\hspace{6pt}\gamma}V_{\gamma\beta}+\partial_{\beta}^{\hspace{6pt}\gamma}V_{\gamma\alpha}\right)$.
In addition, the usual (vector) gauge potential is given by $v^{a}\equiv\left(\gamma^{a}\right)_{\alpha\beta}V^{\alpha\beta}$,
and the (tensor) gauge field strength by $F_{ab}\equiv\partial_{a}v_{b}-\partial_{b}v_{a}=\frac{i}{2}\in_{abc}(\gamma^{c})^{\alpha\beta}f_{\alpha\beta}$.

\section{THE SUPERPROPAGATOR COEFFICIENTS \label{sec:Appedix-B-Prop-Coeff}}

In order to calculate the superpropagators (\ref{propagators}) we
start with the matrices (\ref{eq:kernel-ab}), (\ref{eq:kernel-pi})
and (\ref{eq:kernel-api}) and go through all the operations indicated
in formulae (\ref{eq:invers=0000E3o1})-(\ref{eq:invers=0000E3o4}).
These manipulations involve a lot of algebraic calculation using the
operator algebra presented in Sect. \ref{sec:algebra-operators}.
The complete result is very cumbersome. In the results shown below,
we only kept the terms up to linear dependence in the field component
$\sigma_{2}$, which are enough to discuss the possibility of susy
breakdown \cite{burgess,gallegos}. We will also limit our calculations
of the effective potential to the Landau gauge ($\alpha=0$) and so,
for simplicity, in the calculation of the superpropagators we restrict
ourselves to linear terms in $\alpha$.

The gauge superpropagator $\left\langle AA\right\rangle $ is given
by\begin{equation}
\left\langle T\, A_{\alpha}\left(k,\theta\right)A_{\beta}\left(-k,\theta^{\prime}\right)\right\rangle =i\,\sum\limits _{i=0}^{5}\left(r_{i}\, R_{i,\alpha\beta}+s_{i}\, S_{i,\alpha\beta}+m\, M_{\alpha\beta}+n\, N_{\alpha\beta}\right)\delta^{2}\left(\theta-\theta^{\prime}\right),\label{eq:prop-gauge}\end{equation}
with\begin{eqnarray*}
r_{0} & = & -r_{3}=-\frac{1}{2}r_{4}=\frac{1}{2k^{2}}s_{2}=-\frac{\sigma_{1}\sigma_{2}e^{2}}{4k^{2}\left(k^{2}+m_{1}^{2}\right){}^{2}},\\
r_{2} & = & s_{3}=s_{4}=0,\\
r_{5} & = & -\frac{\xi^{2}\alpha\sigma_{1}\sigma_{2}e^{2}M}{2k^{4}\left(k^{2}+m_{1}^{2}\right)\left(k^{2}+M^{2}\right)},\\
s_{0} & = & -\frac{\alpha}{2k^{2}}-\frac{1}{2\left(k^{2}+m_{1}^{2}\right)},\\
s_{1} & = & s_{5}=\frac{\alpha\sigma_{1}\sigma_{2}e^{2}\left(k^{2}\left(1-\xi^{2}\right)+M^{2}\right)}{2k^{4}\left(k^{2}+m_{1}^{2}\right)\left(k^{2}+M^{2}\right)}-\frac{\sigma_{1}\sigma_{2}e^{2}}{4k^{2}\left(k^{2}+m_{1}^{2}\right){}^{2}}.\\
r_{1} & = & \frac{1}{2k^{2}\left(k^{2}+m_{1}^{2}\right)}-\frac{\alpha\left[\left(M^{2}+m_{1}^{2}\right)k^{2}+\left(k^{2}\right)^{2}+M\left(Mm_{1}^{2}+e^{2}\xi^{2}\sigma_{1}\sigma_{2}\right)\right]}{2\left(k^{2}\right)^{2}\left(k^{2}+M^{2}\right)\left(k^{2}+m_{1}^{2}\right)}\\
n & = & 0\text{ \ \ \ \ \ \ \ \ \ }m\sim O(\alpha^{2}),\end{eqnarray*}
where $m_{1}^{2}=e^{2}\sigma_{1}^{2}/2.$

The scalar superpropagator $\left\langle \Pi\Pi\right\rangle $ exhibits
the following structure\begin{equation}
\left\langle T\,\Pi\left(k,\theta\right)\Pi\left(-k,\theta^{\prime}\right)\right\rangle =i\,\left(\sum\limits _{i=0}^{5}a_{i}\, P_{i}\right)\,\delta^{2}\left(\theta-\theta^{\prime}\right),\label{eq:prop-pi}\end{equation}
where\begin{eqnarray*}
a_{0} & = & \frac{M}{k^{2}+M^{2}}+\frac{\xi^{2}\alpha\sigma_{1}e^{2}\left(k^{2}\left(2M\sigma_{1}+\sigma_{2}\right)-M^{2}\sigma_{2}\right)}{2k^{2}\left(k^{2}+M^{2}\right)^{2}},\\
a_{1} & = & -\frac{1}{k^{2}+M^{2}}+\frac{\xi^{2}\alpha\sigma_{1}e^{2}\left(M\left(M\sigma_{1}+2\sigma_{2}\right)-k^{2}\sigma_{1}\right)}{2k^{2}\left(k^{2}+M^{2}\right)^{2}},\\
a_{2} & = & 2k^{2}a_{5}=-\frac{2\xi^{2}\alpha\sigma_{1}\sigma_{2}e^{2}M}{\left(k^{2}+M^{2}\right)^{2}},\\
a_{3} & = & \frac{1}{2}a_{4}=\frac{\xi^{2}\alpha\sigma_{1}\sigma_{2}e^{2}\left(k^{2}-M^{2}\right)}{2k^{2}\left(k^{2}+M^{2}\right)^{2}}.\end{eqnarray*}
The superpropagator $\left\langle \Pi A\right\rangle $ takes the
form \begin{equation}
\left\langle T\,\Pi\left(k,\,\theta\right)A_{\alpha}\left(-k,\,\theta^{\prime}\right)\right\rangle =i\,\sum\limits _{i=1}^{8}b_{i}T_{\alpha}^{i}\,\delta^{2}\left(\theta-\theta^{\prime}\right),\label{eq:prop-mix}\end{equation}
where\begin{eqnarray*}
b_{1} & = & \frac{\xi e\sigma_{2}M\left[(\alpha+1)\left(k^{2}+M^{2}\right)+2m_{1}^{2}\alpha\xi^{2}\right]}{2\left(k^{2}+m_{1}^{2}\right)\left(k^{2}+M^{2}\right)^{2}},\\
b_{3} & = & b_{2}=-\frac{\xi e\sigma_{2}\left[2k^{4}+2k^{2}\left[\alpha\left(\xi^{2}-1\right)m_{1}^{2}+M^{2}\right]-2\alpha\left(\xi^{2}+1\right)m_{1}^{2}M^{2}\right]}{4k^{2}\left(k^{2}+m_{1}^{2}\right)\left(k^{2}+M^{2}\right)^{2}},\\
b_{4} & = & \frac{\xi e\sigma_{2}M\left[(\alpha-1)k^{2}\left(k^{2}+M^{2}\right)-2\alpha m_{1}^{2}\left[\left(\xi^{2}-1\right)k^{2}-M^{2}\right]\right]}{2k^{4}\left(k^{2}+m_{1}^{2}\right)\left(k^{2}+M^{2}\right)^{2}}\\
b_{5} & = & -\frac{\xi\alpha e\left(M\sigma_{1}+\sigma_{2}\right)}{2k^{2}\left(k^{2}+M^{2}\right)},\\
b_{7} & = & -Mb_{8}=\frac{M\sigma_{2}}{\sigma_{1}}b_{6}=\frac{eM\alpha\xi\sigma_{2}}{2k^{2}\left(k^{2}+M^{2}\right)}.\end{eqnarray*}

These coefficients are not exact, they only exhibit the contributions
up to linear terms in $\alpha$ and in $\sigma_{2}$, what is enough
for our purposes. The exact results are rather cumbersome, even if
their calculation do not present any technical difficulty.

Finally, the superpropagator $\left\langle \Sigma\Sigma\right\rangle $
is given by\begin{equation}
\left\langle T\,\Sigma\left(k,\theta\right)\Sigma\left(-k,\theta^{\prime}\right)\right\rangle =i\,\left[c_{0}P_{0}+c_{1}P_{1}\right]\,\delta^{2}\left(\theta-\theta^{\prime}\right),\label{eq:prop-sig}\end{equation}
where\begin{eqnarray*}
c_{0} & = & \frac{M}{k^{2}+M^{2}},\,\,\,\,\,\,\,\, c_{1}=-\frac{1}{k^{2}+M^{2}}.\end{eqnarray*}

\section{TWO-LOOP CALCULATIONS \label{sec:Appendix-C-Two-Loop-Calculations}}

The supergraphs contributing to the effective potential at two-loop
order, in the vacuum bubble method \cite{jackiw}, are shown in Fig.
\ref{fig:Two-loop}. To study the possibility of supersymmetry breaking
it is enough to calculate the effective potential up to linear order
in the component $\sigma_{2}$ of the classical value of the matter
superfield (\ref{eq:sigma-class}). Only diagrams (a), (c) and (d)
in Fig. \ref{fig:Two-loop} have contributions starting linearly in
$\sigma_{2}$ and also, independent or linear in the gauge parameter
$\alpha$. Using the expressions for the superpropagators and performing
the D-algebra with the help of the SusyMath package \cite{Ferrari},
we get the UV finite results: \begin{eqnarray}
U_{2\left(a\right)} & = & \frac{1}{2}\int\frac{d^{3}k}{\left(2\pi\right)^{3}}\frac{d^{3}q}{\left(2\pi\right)^{3}}\left[\frac{\alpha\xi^{2}\sigma_{1}\sigma_{2}e^{4}M\left[M^{2}-(k+q)^{2}\right]k^{2}}{\left(k^{2}+m_{1}^{2}\right){}^{2}(k+q)^{2}\left(M^{2}+q^{2}\right)\left[(k+q)^{2}+M^{2}\right]^{2}}\right.\nonumber \\
 &  & \left.+\frac{\sigma_{1}\sigma_{2}e^{4}M\left[\alpha\xi^{2}k.q+(k+q)^{2}\right]}{\left(k^{2}+m_{1}^{2}\right){}^{2}(k+q)^{2}\left(M^{2}+q^{2}\right)\left[(k+q)^{2}+M^{2}\right]}\right]\end{eqnarray}

\begin{eqnarray}
U_{2\left(c\right)} & =\alpha\xi\sigma_{1}\sigma_{2}Me^{4} & \int\frac{d^{3}k}{\left(2\pi\right)^{3}}\frac{d^{3}q}{\left(2\pi\right)^{3}}\frac{(k+q)^{2}}{k^{2}\left(k^{2}+M^{2}\right)\left(M^{2}+q^{2}\right)\left[(k+q)^{2}+m_{1}^{2}\right]{}^{2}}\end{eqnarray}

\begin{eqnarray}
U_{2\left(d\right)} & = & \frac{1}{4}\int\frac{d^{3}k}{\left(2\pi\right)^{3}}\frac{d^{3}q}{\left(2\pi\right)^{3}}\frac{1}{\left(k^{2}+m_{1}^{2}\right){}^{2}\left(q^{2}+M^{2}\right)\left[(k+q)^{2}+m_{1}^{2}\right]{}^{2}}\nonumber \\
 &  & \left[\frac{\sigma_{1}\sigma_{2}Me^{4}\left(k^{2}+k\cdot q\right)\left[2(1-2\alpha)k^{4}-4\alpha m_{1}^{2}k^{2}-2\alpha m_{1}^{4}\right]}{2k^{4}}\right.\nonumber \\
 &  & \left.-\frac{\alpha\sigma_{1}\sigma_{2}Mm_{1}^{4}e^{4}\left(k^{2}+k\cdot q\right)}{(k+q)^{4}}-\frac{\sigma_{1}\sigma_{2}Mm_{1}^{2}e^{4}\left(2\alpha k^{2}+m_{1}^{2}\right)\left(k^{2}+k\cdot q\right)}{k^{2}(k+q)^{2}}\right]\end{eqnarray}

The other supergraph contributions are of order $\alpha^{2}$ or $\sigma_{2}^{2}$,
that is, $U_{2\left(b\right)}=\mathcal{O}\left(\alpha^{2},\sigma_{2}\right),$
$U_{2\left(e\right)}=\mathcal{O}\left(\alpha^{2},\sigma_{2}^{2}\right)$,
$U_{2\left(f\right)}=U_{2\left(g\right)}=\mathcal{O}\left(\alpha,\sigma_{2}^{2}\right)$.

\end{document}